\documentclass[prd,aps,preprint,tightenlines,showpacs,nofootinbib,superscriptaddress]{revtex4-1}
\usepackage{mathrsfs}
\usepackage{amsfonts}
\usepackage{amsmath}
\usepackage{array}
\usepackage{verbatim}
\usepackage{bm}
\usepackage{epsfig}
\usepackage{relsize}
\usepackage{lineno}
\usepackage{float}
\usepackage{multirow}
\RequirePackage{xspace}

\def\lsim{\raise0.3ex\hbox{$<$\kern-0.75em\raise-1.1ex\hbox{$\sim$}}}
\def\gsim{\raise0.3ex\hbox{$>$\kern-0.75em\raise-1.1ex\hbox{$\sim$}}}

\def\pom{{I\!\!P}}

\def\beq{\begin{equation}}
\def\eeq{\end{equation}}
\def\bea{\begin{eqnarray}}
\def\eea{\end{eqnarray}}
\def\bq{\begin{quote}}
\def\eq{\end{quote}}

\def\gappeq{\mathrel{\rlap {\raise.5ex\hbox{$>$}}
{\lower.5ex\hbox{$\sim$}}}}

\def\lappeq{\mathrel{\rlap{\raise.5ex\hbox{$<$}}
{\lower.5ex\hbox{$\sim$}}}}

\def\Toprel#1\over#2{\mathrel{\mathop{#2}\limits^{#1}}}

\def\pom{{I\!\!P}}

\begin{document}


\title{Probing hidden - bottom pentaquarks in fixed - target collisions at the LHC}

\author{Ya-Ping Xie}
\email{xieyaping@impcas.ac.cn}
\affiliation{Institute of Modern Physics, Chinese Academy of Sciences,
Lanzhou 730000, China}
\affiliation{University of Chinese Academy of Sciences, Beijing 100049, China}

\author{V.~P. Gon\c{c}alves}
\email{barros@ufpel.edu.br}
\affiliation{High and Medium Energy Group, \\
Instituto de F\'{\i}sica e Matem\'atica, Universidade Federal de Pelotas\\
Caixa Postal 354, CEP 96010-900, Pelotas, RS, Brazil}
\affiliation{Institute of Modern Physics, Chinese Academy of Sciences,
Lanzhou 730000, China}
\date{\today}

\begin{abstract}
In this paper we investigate the possibility of searching for  the hidden - bottom pentaquark states in photon -- induced interactions at the LHC. 
We consider the presence of the $P_b$ resonance  in the $s$ -- channel of the $\gamma p \rightarrow \Upsilon p$ reaction and estimate its contribution for
the exclusive $\Upsilon$ photoproduction in the fixed - target mode of the LHC.
Predictions for the total cross sections, rapidity and transverse momentum distributions are derived using the STARlight Monte Carlo considering  $Pb - p$, $Pb - He$ and $Pb - Ar$ fixed - target collisions at the LHC. Our results indicate that the presence of the $P_b$ resonance implies an enhancement in the rapidity distribution in the kinematical range covered by the LHCb detector. We  demonstrate that  the $P_b$ contribution for the $\Upsilon$ photoproduction becomes dominant if  kinematical cuts are imposed on the rapidity and transverse momentum of the final state. These results indicate that an  experimental analysis of the $\Upsilon$ photoproduction in fixed -- target collisions  can provide complementary and independent checks of the existence of  these states, and help to understand their underlying nature.
\end{abstract}

\maketitle

\section{Introduction}
\label{intro}

In the last  years many exotic hadrons, which are states which does not appear to fit with the
expectations for an ordinary $q\bar{q}$ or $qqq$ hadrons in the quark model, have been observed at various experimental facilities (For reviews see, {\it e.g.}, Refs. 
\cite{Olsen:2017bmm,Liu:2019zoy}). Such results have motivated a series of studies focused on the description of the internal structure of the exotic hadrons as well as the proposition of new channels to search and constrain the properties of these states. 
In particular, the study of hidden - heavy quark pentaquark states has received a lot of attention (See e.g. Refs.  
\cite{Chen:2019asm,Xiao:2019mst,Chen:2019bip,Cheng:2019obk,Liu:2019tjn,He:2019ify,Cao:2019kst,Wang:2019krd,Wang:2019zaw,Xie:2020wfe,vicmiguel2,Guo:2019kdc,Xiao:2020frg,Xie:2020niw}), strongly motivated by the results reported by the 
LHCb Collaboration for the $\Lambda_b^0 \rightarrow J/\Psi p K^-$ decay, which indicate the existence of three narrow pentaquark states: $P_c(4312)$, $P_c(4440)$ and $P_c(4457)$ \cite{lhcb_penta}. Such observation was confirmed by the 
D0 Collaboration, which have analyzed the production of these states in $p\bar{p}$ collisions \cite{d0_penta} and found  an enhancement in the $J/\Psi p$ invariant mass consistent with a sum of the resonances $P_c(4440)$ and $P_c(4457)$ reported by the LHCb Collaboration. On the other hand, 
the GlueX Collaboration at the Thomas  Jefferson National Accelerator Facility (JLab) did not see evidence for them \cite{gluex_penta}, which implies an upper limit on the branching ratios  ${\cal{B}}(P_c \rightarrow J/\Psi p)$. In addition, the nature of these states is still a theme of intense debate. One of the most promising alternatives
to probe  the existence  and to decipher the nature of these states is the study of the photoproduction of the $P_c$ state in the $\gamma p \rightarrow J/\Psi p$ reaction at low center -- of - mass energies ($\sqrt{s} \le 20$ GeV) \cite{Xie:2020niw} , which can be realized experimentally at JLab and in the future electron - ion collider in China (EicC) \cite{EicC} \footnote{For large values of $\sqrt{s}$, the $P_c$ states are produced at very forward rapidities, beyond the rapidity range covered by the detectors proposed to be installed in the EIC at US and LHeC at CERN.}. Another possibility is to study the 
 exclusive $J/\Psi$ photoproduction in fixed - target  collisions at the LHC \cite{vicmiguel}. In recent years, the study of fixed - target collisions at the LHC became a reality \cite{lhcfixed} and an comprehensive fixed - target program using the LHC beams is expected to be developed in forthcoming years \cite{after}. As shown in Ref. \cite{vicmiguel}, the fixed - target collisions allows to probe the photon - induced interactions in a  limited energy range, dominated by low - energy interactions, which is the region of interest for the study of the $P_c$ production. Such aspect was explored in Ref. \cite{vicmiguel2}, which have  investigated the impact of the $P_c$ resonances on the $J/\Psi$ photoproduction at the LHC and demonstrated that  the presence of the resonances modifies the associated rapidity distribution in the rapidity range probed by the LHCb detector. As a consequence, future experimental analysis of fixed -- target collisions are expected to improve our understanding of the properties of the  $P_c$ states.

\begin{figure}[t]
\includegraphics[scale=0.5]{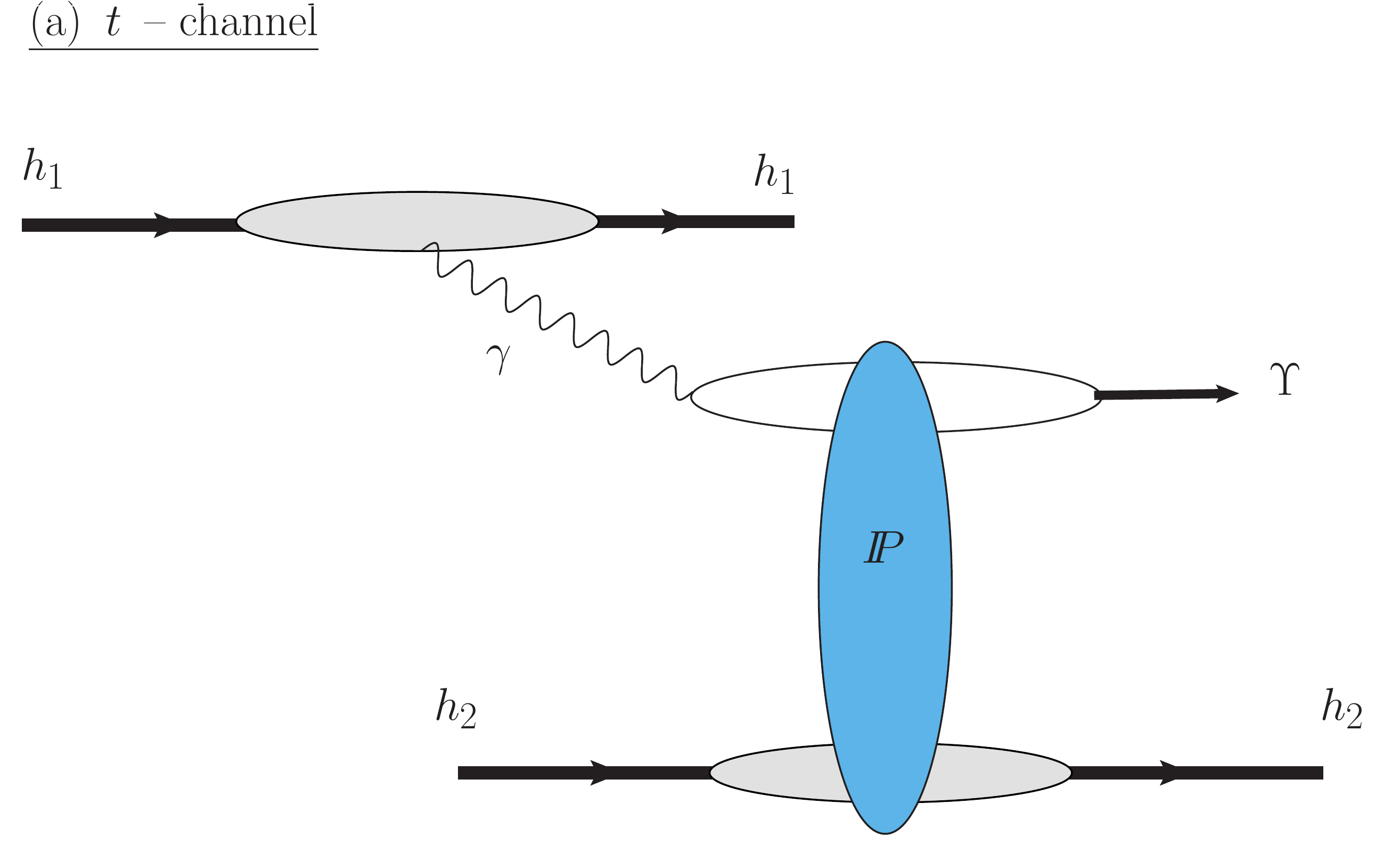} \\
\includegraphics[scale=0.5]{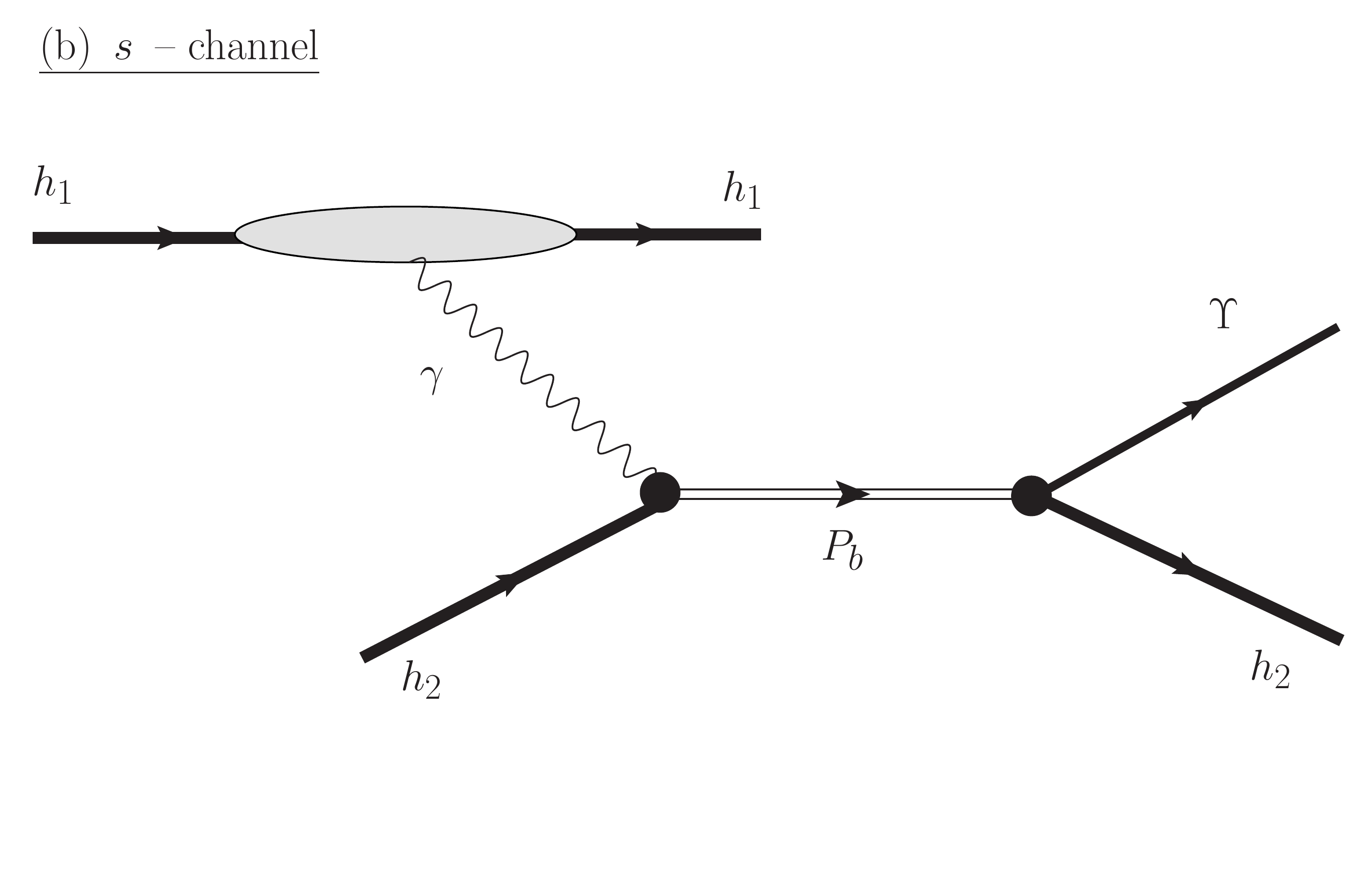} 
\caption{ The exclusive $\Upsilon$ photoproduction in hadronic collisions process through  (a) the Pomeron exchange in the $t$ -- channel and (b) the $P_b$ production in the $s$ -- channel. }
\label{Fig:diagrama}
\end{figure}

In this paper we will extend the analysis performed in Ref. \cite{vicmiguel2} for the case of hidden -- bottom pentaquark states, denoted $P_b$ hereafter, which can contribute for the exclusive $\Upsilon$ photoproduction in fixed - target collisions at the LHC. Such states were not yet observed and are expected to be formed by three light quarks and a bottom quark pair. Distinctly from the $P_c$ states, the $P_b$ states cannot be produced through the decay of heavier baryons, which motivates the investigation of its direct production in hadronic and photon -- induced interactions (See e.g. Refs.   \cite{Wang:2019zaw,Cao:2019gqo}). 
We will focus on ultraperipheral heavy -- ion collisions \cite{upc}, which are characterized by large impact parameter $b > R_{h_1} + R_{h_2}$ ($R_{h_i}$ is the hadron radius). For this case, the photon -- induced interactions become dominant and hadronic collisions can be used to study the production of the $P_b$ states in the $\gamma h \rightarrow \Upsilon h$ reaction. 
 In our analysis we will extend  previous studies for the exclusive $\Upsilon$ photoproduction in hadronic collisions at the LHC,  which  estimated the contribution of the Pomeron exchange in the $t$ -- channel for the cross section [represented in Fig. \ref{Fig:diagrama} (a)], by  taking into account  the  contribution associated to the presence of the $P_b$ resonances in the $s$ -- channel [See Fig. \ref{Fig:diagrama} (b)].  
In our analysis the contribution associated to the Pomeron exchange will be estimated using the  STARlight Monte Carlo \cite{starlight}, which successfully describe the main aspects of the photon - induced interactions at the LHC (See e.g. Refs. \cite{alice,alice2,alice3,lhcb,lhcb2,lhcb3}). 
On the other hand, we will estimate the $s$ -- channel contribution using the approach proposed in Ref. \cite{Wang:2019zaw}, which we have implemented in the STARLight MC. The exclusive $\Upsilon$ photoproduction cross section will be estimated considering the contribution of both channels for $Pb - p$, $Ar - p$, $Pb - Ar$  and $Pb - He$ fixed target collisions at the LHC and our predictions for the rapidity and transverse momentum distributions will be presented. As we will show below, our results indicate that  the study of the  $\Upsilon$ photoproduction in fixed -- target collisions can be  useful to probe the existence of the $P_b$ pentaquark states.  

The paper is organized as follows. In the next Section a brief review of formalism needed to describe the $\Upsilon$ photoproduction via the diagrams presented in Fig. \ref{Fig:diagrama}. In Section \ref{res}, we will present our predictions for the total cross sections and associated rapidity and transverse distributions considering a projectile $Pb$ beam and different nuclei as fixed target. Finally, in Section \ref{conc} we will summarize our main results and conclusions.

\section{Formalism}
\label{form}

In what follows we will present a brief review of the formalism needed to describe the exclusive $\Upsilon$ photoproduction  in ultraperipheral collisions (UPCs). In UPCs,   the photon -- induced interactions become dominant and the hadronic cross sections  can be factorized in terms of the equivalent flux of photons of the incident hadrons and the photon-photon or photon-target  cross section. We will focus on photon -- hadron interactions, where the  photon stemming from the electromagnetic field of one of the two  hadrons  interact directly with the other hadron \cite{upc}.  In this case, the cross section for the exclusive $\Upsilon$ photoproduction can be expressed by,
\begin{widetext}
\begin{equation}
   \sigma(h_1 + h_2 \rightarrow h_1 \, V \, h_2;\,s) =  \int d\omega \,\, n_{h_1}(\omega) \, \sigma_{\gamma h_2 \rightarrow V \, h_2}\left(W_{\gamma h_2}  \right) + \int d \omega \,\, n_{h_2}(\omega)
   \, \sigma_{\gamma h_1 \rightarrow V \, h_1}\left(W_{\gamma h_1}  \right)\,  \; , 
\label{eq:sigma_pp}
\end{equation}
\end{widetext}
where $\sqrt{s}$ is center-of-mass energy for the $h_1 h_2$ collision ($h_i$ = p,A),  $\omega$ is the energy of the photon emitted by the hadron and $n_h$ is the equivalent photon flux of the hadron $h$ integrated over the impact parameter. Moreover, $\sigma_{\gamma h \rightarrow V \, h}$ describes the vector meson production in photon - hadron interactions, which will be given in our analysis by the sum of the contributions associated to the Pomeron exchange and $P_b$ resonances.
Following Ref. \cite{starlight}, the Pomeron contribution for photon -- proton interactions will be described by a parameterization inspired in the Regge theory given by
\begin{eqnarray}
\sigma^{\pom}_{\gamma p \rightarrow \Upsilon  p} = \sigma_{\pom} \times W_{\gamma p}^{\epsilon} \times \, \left(1 - \frac{(m_p + m_{\Upsilon})^2}{W_{\gamma p}^2}\right)^2 \,\,,
\label{sig_gamp_pom}
\end{eqnarray}
with $\sigma_{\pom} = 6.4$ pb and $\epsilon = 0.74$ being derived fitting the HERA data \cite{hera} and the term in parenthesis describes the behaviour of the cross section for energies near to the  threshold of production.
On the other hand, in order to estimate the contribution associated to the $P_b$ resonance, denoted by  $\sigma^{P_c}_{\gamma p \rightarrow \Upsilon  p}$, we follow the approach proposed in  Ref. \cite{Wang:2019zaw},
 where the photoproduction of the $P_b$ states is estimated within the framework of an  effective Lagrangian approach combined with the vector meson dominance assumption \cite{sakurai}. In this approach, the Lagrangians needed to estimated to estimate the photon -- nucleon cross section are expressed by
\begin{eqnarray}
{\cal{L}}_{\gamma N P_b} = \frac{eh}{2m_p} \bar{N} \sigma_{\mu \nu} \partial^{\mu} A^{\mu} \, P_b + \,\,{\mbox{h.c.}} \,\,\\
{\cal{L}}_{P_b \Upsilon N} = g_{P_b \Upsilon N} \bar{N} \gamma_5 \gamma_{\mu} P_b\Upsilon^{\mu} + \,\,{\mbox{h.c.}}
\end{eqnarray} 
where $N$, $A$, $P_b$ and $\Upsilon$ characterize the nucleon, photon, $P_b$ state and $\Upsilon$ meson fields, respectively. Moreover,  $g_{P_b \Upsilon N}$ is the $P_b - \Upsilon - N$ coupling constant, which can be estimated from the  $\Gamma_{P_b \rightarrow \Upsilon N} $ decay width, and $eh$ is the electromagnetic coupling related to $\gamma N P_b$ vertex.  As in Refs. \cite{Wang:2019zaw,Xie:2020wfe}, we will assume that  the $P_b$ state is characterized by $J^P = {\frac{1}{2}}^-$, $M = 11080$ MeV and  $\Gamma_{P_b \rightarrow \Upsilon N} = 0.38$ MeV. In addition, we assume  $g_{P_b \Upsilon N} = 0.074$ and $eh = 0.00016$. The resulting  $\sigma^{P_c}_{\gamma p \rightarrow \Upsilon  p}$ cross section has been implemented in the STARLight MC. 
Finally, for a nuclear target, we will disregard the nuclear effects  and assume  that $\sigma_{\gamma A \rightarrow \Upsilon A} = A \times \sigma_{\gamma p\rightarrow \Upsilon p}$. One has that for the light targets considered in this paper, this is a reasonable approximation, since the $\Upsilon$ state is a  compact object, with negligible interactions with the nuclear medium.

In order to estimate the cross section, Eq. (\ref{eq:sigma_pp}), we must specify the photon spectrum associated to protons and nuclei. In our analysis, we use the approach proposed in Refs. \cite{klein,starlight}, which is implemented in the  STARlight MC. In this approach, the photon spectrum is calculated as follows 
\begin{eqnarray}
n(\omega) = \int \mbox{d}^{2} {\mathbf b} \, P_{NH} ({\mathbf b}) \,   N\left(\omega,{\mathbf b}\right) \,\,, 
\end{eqnarray}
where $P_{NH} ({\mathbf b})$ is the probability of not having a hadronic interaction at impact parameter ${\mathbf b}$  and the number of photons per unit area, per unit energy, derived assuming a point-like form factor, is given by 
\begin{equation}
N(\omega,{\mathbf b}) = \frac{Z^{2}\alpha_{em}}{\pi^2} \frac{\omega}{\gamma^{2}}
\left[K_1^2\,({\zeta}) + \frac{1}{\gamma^{2}} \, K_0^2({\zeta}) \right]\,
\label{fluxo}
\end{equation}
where $\zeta \equiv \omega b/\gamma$ and $K_0(\zeta)$ and  $K_1(\zeta)$ are the
modified Bessel functions. For nuclear collisions, one has that $P_{NH} ({\mathbf b}) = \exp[-\sigma_{NN} \, T_{AA} ({\mathbf b})]$, where $\sigma_{NN}$ is the nucleon - nucleon interaction cross section and $T_{AA} ({\mathbf b})$ is the nuclear overlap function, which is assumed to be the convolution of a hard sphere potential with a Yukawa potential of range 0.7 fm.
On the other hand, for proton - nucleus collisions, $P_{NH} ({\mathbf b}) = \exp[-\sigma_{NN} \, T_{A} ({\mathbf b})]$, where $T_{A} ({\mathbf b})$ is the nuclear thickness function (For details see Ref. \cite{starlight}).

\section{Results}
\label{res}

In what follows we will present our estimates for the exclusive $\Upsilon$ photoproduction in fixed - target $Pb - A$ collisions at $\sqrt{s} = 110$ GeV and 69 GeV, respectively, assuming $A = p, He, Ar$.  For comparison, we also will present results for $Ar - p$ collisions.  We will restrict our analysis to the $\Upsilon(1S)$ state, but the results can be easily extended for excited states.  Moreover, we will consider the full LHC kinematical range as well as the kinematical range probed by the LHCb detector. In the latter case, we select the events in which the  vector meson is produced in the rapidity range $2 \le y \le 4.5$.

\begin{figure}[H]
\begin{center}
\includegraphics[width=0.55\textwidth]{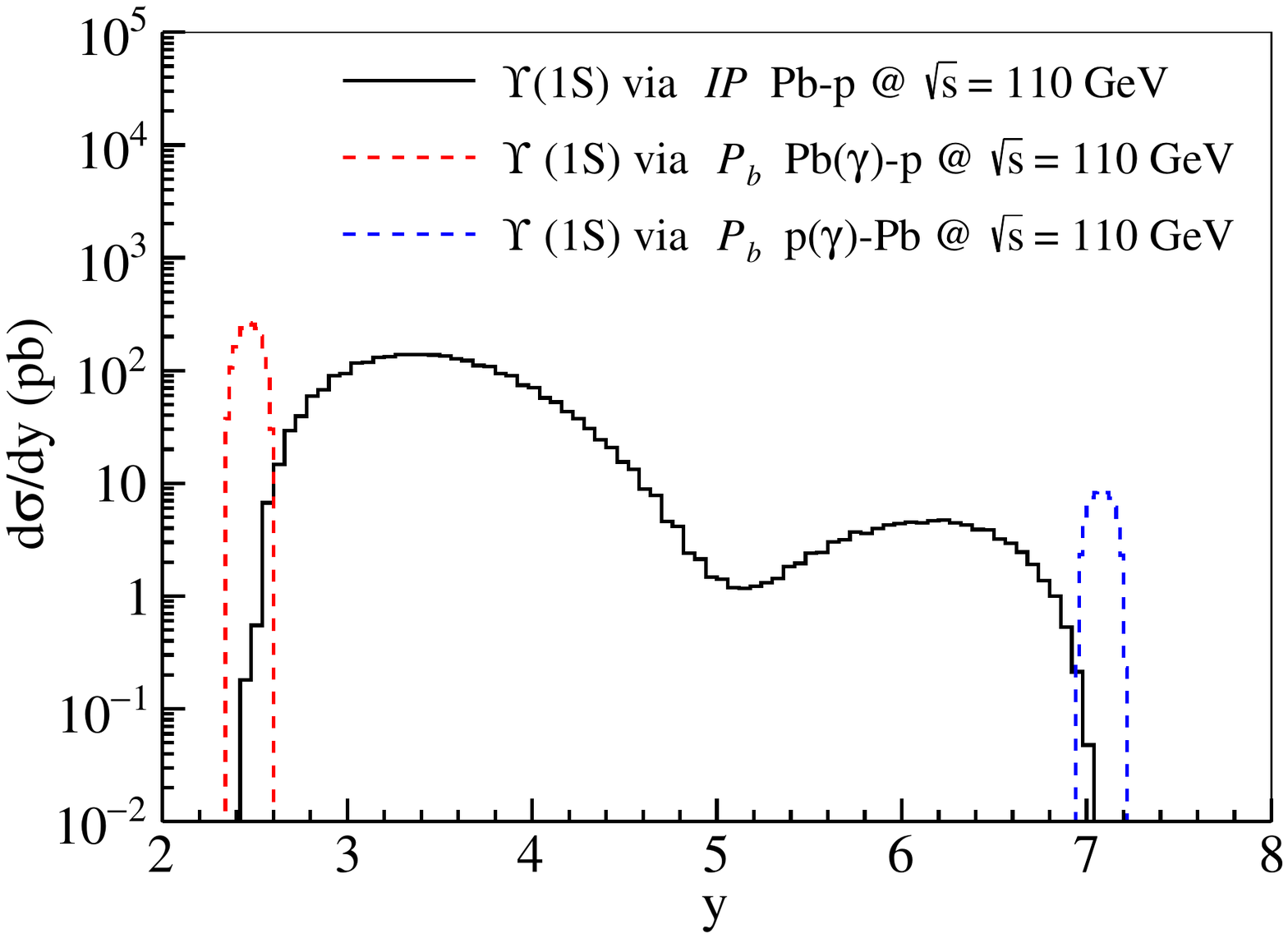} \\  
\includegraphics[width=0.55\textwidth]{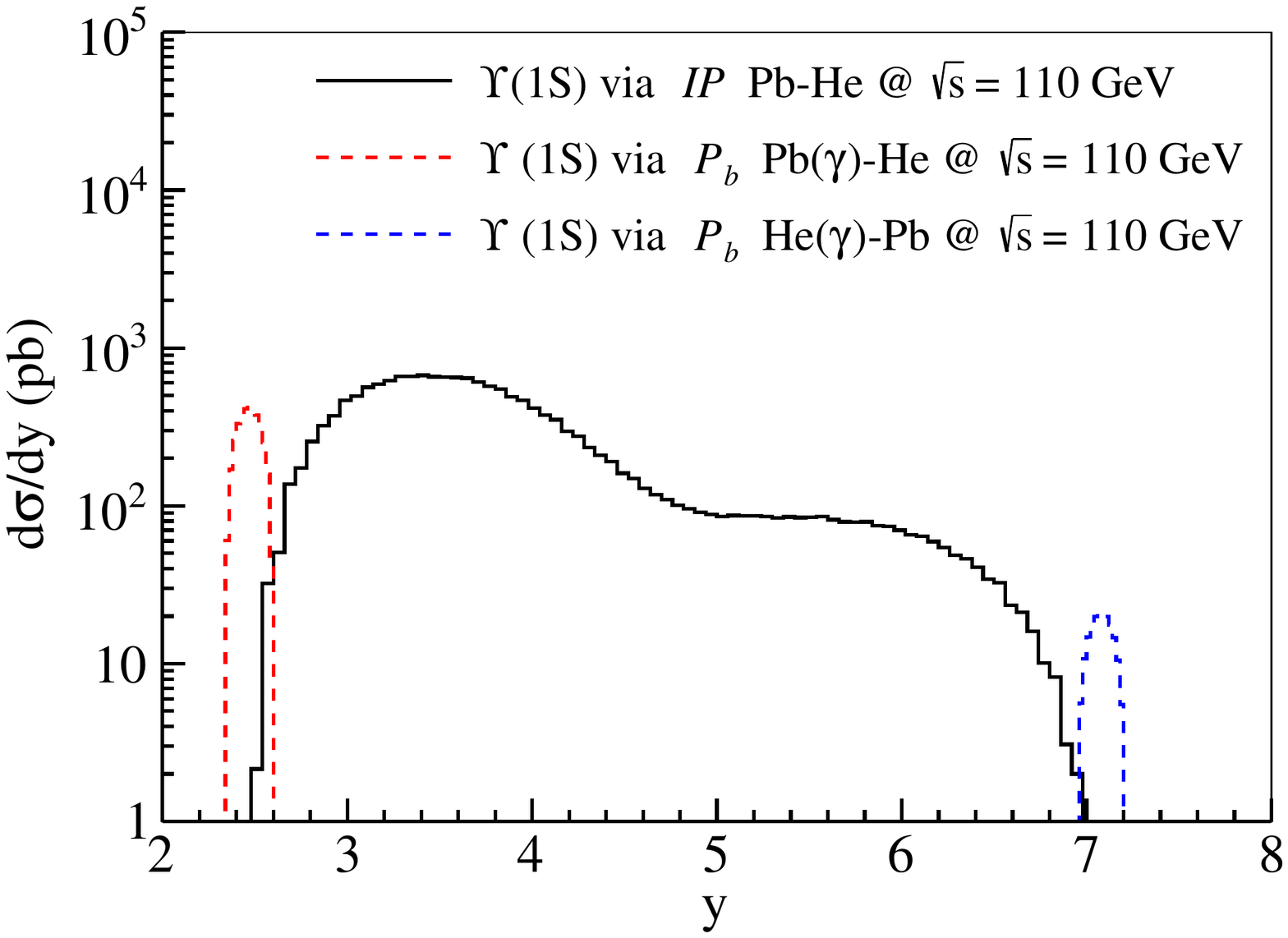}  \\
\includegraphics[width=0.55\textwidth]{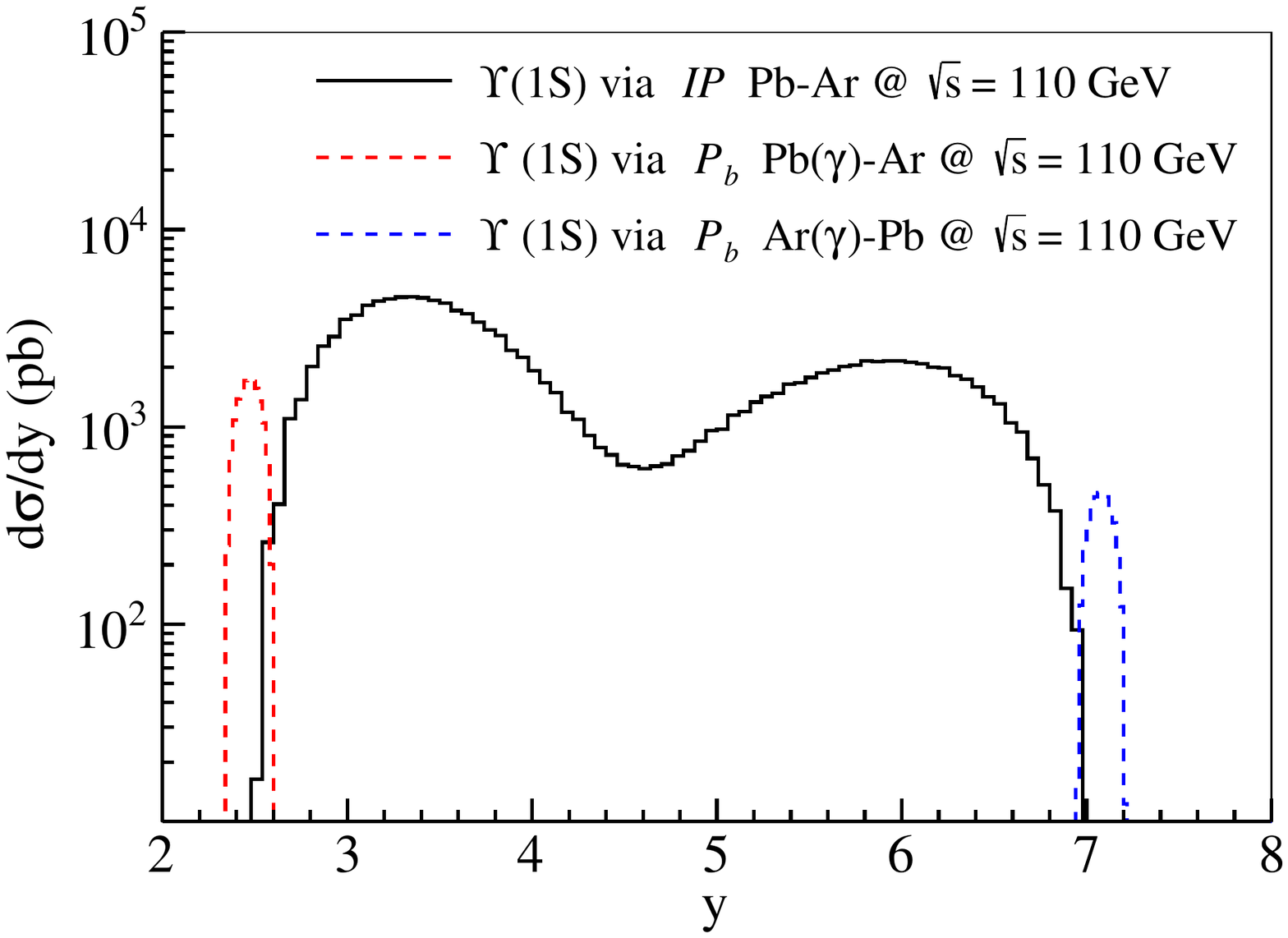}

\end{center}
\caption{Rapidity distributions for the exclusive $\Upsilon$ photoproduction in $Pb - p$ (upper panel), $Pb - He$ (center panel) and and $Pb - Ar$ (lower panel) fixed target collisions at $\sqrt{s} = 110$ GeV.}
\label{Fig:rap}
\end{figure}

Initially, in Fig. \ref{Fig:rap}, we present our predictions for the rapidity distributions considering distinct projectile - target configurations and $\sqrt{s} = 110$ GeV. The contributions associated to the Pomeron and $P_b$ resonance are presented separately. As we are considering the collision of non - identical hadrons, the magnitude of the photon fluxes associated to the two incident hadrons are different, which implies asymmetric rapidity distributions. Moreover, one has that distinctly from the predictions for the collider mode presented e.g. in Ref. \cite{run2}, the maximum of the distributions occur in fixed - target collisions for forward rapidities.  In particular, it occurs in the kinematical range probed by the LHCb detector. One has that the contribution of  the $P_b$ resonance implies the presence of two peaks in the rapidity distribution. Such result is expected. For a fixed value of rapidity $y$, such distribution is determined  by the sum of two terms, the first one being  determined by the $\gamma h$ cross section for  $W_{\gamma h_2}^2 \propto e^{+y}$, while the second one probes the cross section for $W_{\gamma h_1}^2 \propto e^{-y}$ [See Eq. (\ref{eq:sigma_pp})]. Therefore, the resonance condition, $W_{\gamma h} \approx M$, is satisfied for two distinct values of rapidity. Moreover, we have that the magnitude of the peak is larger when the photon is emitted by the $Pb$ beam, which is also expected, since the photon flux is proportional to $Z^2$. One very important aspect, is that the highest peak is predicted to occur in the range probed by the LHCb detector, with its contribution for the $\Upsilon$ photoproduction being larger than that associated to the Pomeron exchange. Therefore, our results indicate that a future experimental analysis of fixed - target collisions by the LHCb detector can be very useful to probe the existence and properties of the $P_b$ resonance.

\begin{figure}[t]
\includegraphics[scale=0.5]{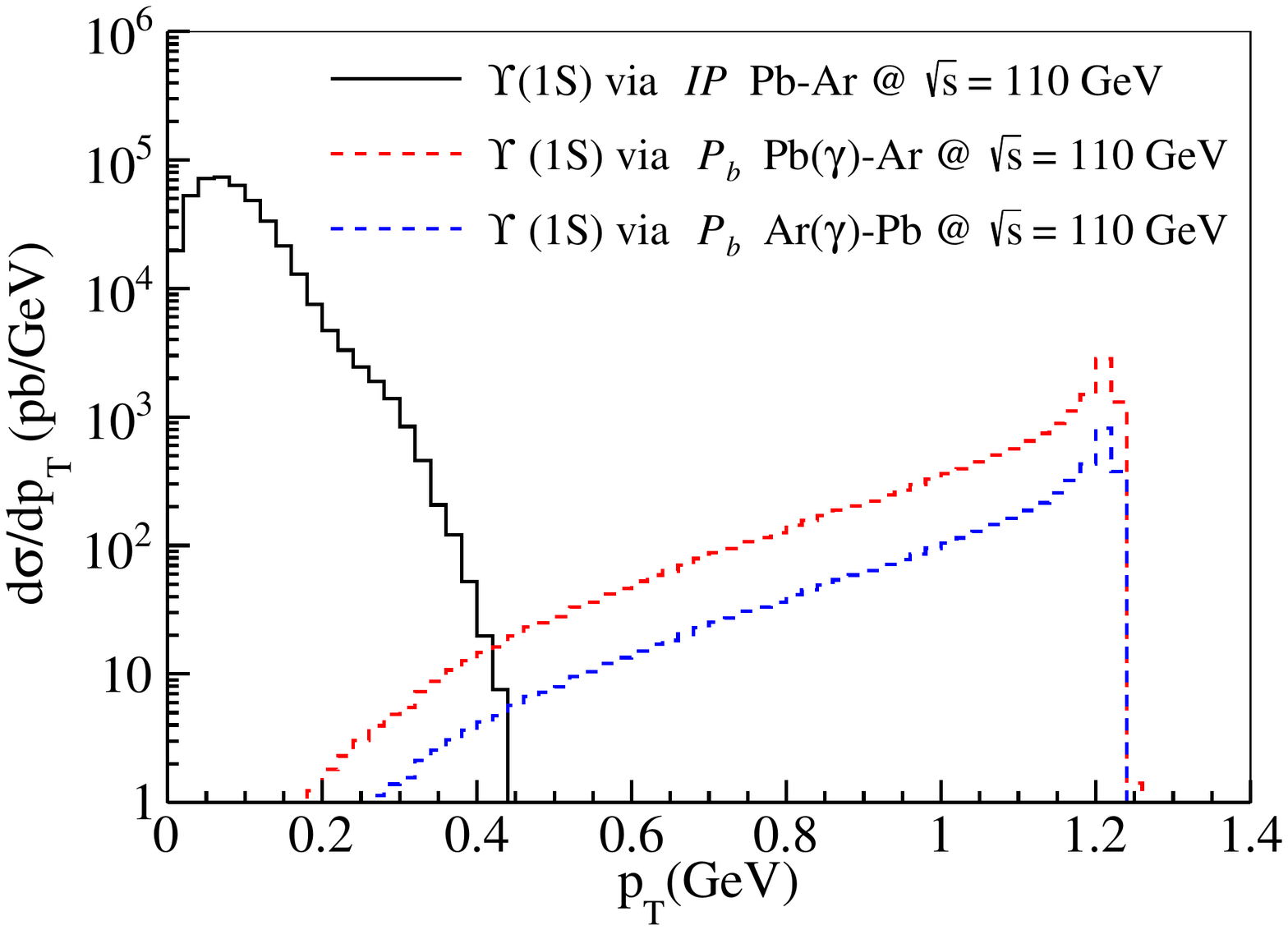} 
\caption{ The transverse momentum distribution for the exclusive $\Upsilon$ photoproduction in $Pb - Ar$ fixed target collisions at $\sqrt{s} = 110$ GeV.  }
\label{Fig:pt}
\end{figure}

In Fig. \ref{Fig:pt} we present our predictions for the transverse momentum distribution
associated to the exclusive $\Upsilon$ photoproduction in $PbAr$ collisions at $\sqrt{s} = 110$ GeV. Similar results are obtained for other configurations of projectile and target. The predictions for the Pomeron and $Pb$ resonance are presented separately. For the Pomeron contribution, one has that $p_T$ spectrum of the vector meson is determined by the sum of the photon momentum with the exchanged momentum in the interaction between the vector meson and the target. While the photon momentum is defined by the equivalent photon approximation, the exchanged one is determined  by the form factor of the target.  As both the photon and scattering transverse momenta are small, we expect that the meson $p_T$ spectrum  associated to the Pomeron exchange will be dominated by small values of transverse momentum, being strongly suppressed at large $p_T^2$. This behaviour is observed in Fig. \ref{Fig:pt}. On the other hand, the spectrum associated to the $P_b$ contribution is characterized by vector mesons with larger transverse momentum. Such result indicates that the Pomeron contribution can be strongly suppressed selecting the events with $p_T \ge 0.4$ GeV.

Finally, our predictions for the total cross sections are presented in Table \ref{table:XSeC} considering different rapidity ranges and $\sqrt{s} = 110$ and 69 GeV, with the predictions for $\sqrt{s} = 69$ GeV being presented in parenthesis. As expected, the Pomeron contribution is strongly energy dependent, since $\sigma^{\pom} \propto W_{\gamma p}^{\epsilon}$. Moreover, one has that the cross section increases with atomic number of the particles that interact due to the factor  $Z^2$ factor present in the nuclear photon flux. In agreement with the results presented in Fig. \ref{Fig:rap}, the selection of the LHCb rapidity range reduces the magnitude of the cross sections, being this suppression larger for the Pomeron contribution in $Pb - Ar$ collisions. Our results indicate that the  $P_b$ contribution is not negligible for the $\Upsilon$ photoproduction in the kinematical range covered by the LHCb detector. It is important to emphasize that this contribution becomes dominant if we select the events in the rapidity range $ 2.0 \le y \le 2.7 $, as demonstrated in the last two lines of Table \ref{table:XSeC}. 
Such results indicate that  the separation of the $P_b$ contribution in a future experimental analysis is, in principle, feasible.

\begin{widetext}
\begin{center}
\begin{table}[t]
\begin{tabular}{|c|c|c|c|c|c|c|}
\hline 
 & {\bf Pb - p}  & {\bf Pb - He} & {\bf Pb - Ar} & {\bf Ar - p}\tabularnewline
\hline 
\hline 
$\sigma^{\pom}$ (Full rapidity range) & 168.0 (13.0)  & 1000.0 (140.0)  & 8400.0 (870.0)  & 22.0 (2.9)\tabularnewline
\hline 
$\sigma^{P_b}$ (Full rapidity range)  & 46.0 (12.0)  & 75.0 (24.0)    & 380.0 (80.0)  & 3.6 (1.3) \tabularnewline
\hline 
\hline
$\sigma^{\pom}$ ($ 2.0 \le y \le 4.5 $) & 160.0 (12.0)  & 860.0 (100.0)  & 5100.0 (370.0)  & 18.0 (2.6)\tabularnewline
\hline 
$\sigma^{P_b}$ ($ 2.0 \le y \le 4.5 $)  & 45.0 (11.0)  & 72.0 (22.0)    & 300.0 (58.0)  & 3.3 (1.2) \tabularnewline
\hline
\hline 
$\sigma^{\pom}$ ($ 2.0 \le y \le 2.7 $) & 2.2 (0.52)  & 10.0 (2.7)  & 78.0 (16.0)  & 0.16 (0.058)\tabularnewline
\hline 
$\sigma^{P_b}$ ($ 2.0 \le y \le 2.7 $) & 45.0 (11.0)  & 72.0 (22.0)    & 300.0 (58.0)  & 3.3 (1.2) \tabularnewline
\hline 
\hline 
\end{tabular}
\caption{Total cross sections (in pb) for the exclusive $\Upsilon$ photoproduction in fixed - target collisions at the LHC considering different rapidity ranges and  $\sqrt{s} = 110 \,(69)$ GeV. }
\label{table:XSeC}
\end{table}
\end{center}
\end{widetext}

\section{Summary}
\label{conc}

In recent years, the study of photon - induced interactions in hadronic colliders has allow us to improve our understanding about the partonic structure of hadrons. In this paper, we have proposed to search for the $P_b$ resonance in the exclusive $\Upsilon$ photoproduction considering fixed - target collisions at the LHC.  Our analysis is strongly motivated by  the proposition of the AFTER@LHC experiment and by the study of beam - gas interactions, recently performed by the LHCb detector, as well by recent theoretical studies that demonstrated that the study of the exclusive vector meson photoproduction is feasible in the fixed - target mode. 
 We have estimated the total cross sections, rapidity and transverse momentum distributions for the exclusive $\Upsilon$  photoproduction considering different configurations of projectile and target and assuming  $\sqrt{s} = 110$ and $69$ GeV. The Pomeron and $P_b$ resonance contributions were estimated using the STARlight Monte Carlo, which allowed to take into account some typical LHCb requirements for the selection of exclusive events. 
Our goal was to verify if the study of this process can be useful to probe  the existence of the $P_b$ resonance. We shown that the presence of the resonance modifies the associated rapidity distribution due to the large enhancement of the $\gamma p \rightarrow \Upsilon p$ cross section near the threshold. We have demonstrated that the rapidity distribution is enhanced in rapidity range covered by the LHCb detector, and that events associated to the $P_b$ contribution can be separated by imposing additional cuts on the rapidity range and transverse momentum of the vector meson. Finally, our results indicate that the study of the  exclusive $\Upsilon$  photoproduction in fixed - target collisions at the LHC provide a complementary and independent probe of the hidden - bottom  pentaquark states, and a future experimental analysis will be useful  to understand their underlying nature.


\section*{Acknowledgments}
The work is partially supported by the Strategic Priority Research Program of Chinese Academy of Sciences (Grant NO. XDB34030301). 
VPG was  partially financed by the Brazilian funding
agencies CNPq,   FAPERGS and  INCT-FNA (process number 
464898/2014-5).



\end{document}